# Creative NFT-Copyrighted AR Face Mask Authoring Using Unity3D Editor

Mohamed Al Hamzy*, Shijin Zhang$, Hong Huang*, Wanwan Li*

University of South Florida*   Columbia University$

**Abstract:** In this paper, we extend well-designed 3D face masks into AR face masks and demonstrate the possibility of transforming this into an NFT-copyrighted AR face mask that helps authenticate the ownership of the AR mask user so as to improve creative control, brand identification, and ID protection. The output of this project will not only potentially validate the value of the NFT technology but also explore how to combine the NFT technology with the AR technology so as to be applied to the e-commerce and e-business aspects of the multimedia industry.

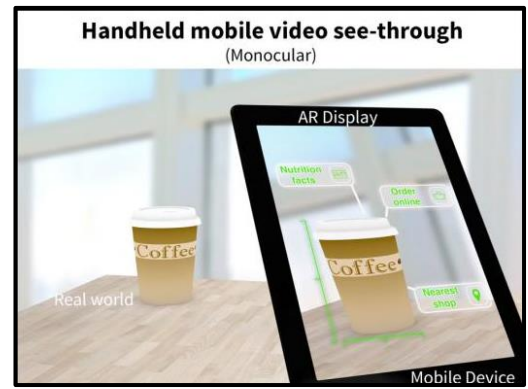

Figure 1: How augmented reality is extended real-time to a mobile device.

## Introduction

As the promising trend of the digital transformation of e-commerce and e-business, VR and AR extended interactive interfaces have gained an increasing amount of attention from researchers. However, given the current stage of digital transformation technologies, combining the VR/AR interface with the NFT-Copyright authentication is still a novel area where there is a lack of experiment results and technical approaches as support. Especially, for the area of the e-commerce and e-business aspects of the multimedia industry. As multimedia technologies become democratic, more and more self-media hosts and internet celebrities are emerging. This fact is leading to a new trend in online traffic-driven e-commerce and e-business which is based on selling and purchasing the self-media platforms' reputation. Choosing a unique AR face mask for an anonymous self-media channel became a trendy topic for online hosts. There are several aspects that need to be considered during the AR face mask decision-making: (1) the AR face mask must be unique, innovative, and attractive; (2) the AR face mask must be copyrighted to avoid the coping and reusing by other self-media hosts. All of these observations show demand for copyrighted AR face masks.

On the other hand, Augmented Reality is one of the latest additions to that. It has been relevant since the 1950s. Lately, and more than ever, AR technology has enabled immersive experiences in the environment when outputted on a real stream. Unity Technologies, a leading game engine and platform, has been responsible for significant advancements in augmented reality (AR) in recent years. The company had been investing in developing tools to support AR projects, allowing developers to easily create interactive AR experiences. From ARKit and ARCore integration to AR Foundation and the Multi-platform AR SDK, Unity has made it easier for developers to reach a wider audience and bring their AR visions to life. These advancements have helped progress the AR industry.

An AR system can be simplified to be of three components: Recognition, tracking and mixing. These are considered to be the progressive steps that happen in real-time [1]. A virtual object is typically imposed on the recognized surface captured by a lens. Visually captured snapshots of the target object are sent to the computer. Which then interprets the visual space and places a graphical object on top of the desired marker. The results of this are viewable on the interface, typically inputted from the lens of a camera.

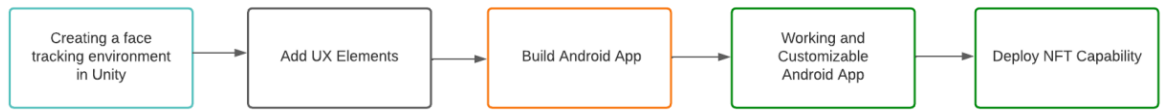

Figure 2: Process Flow of AR mask application

Unity3D uses ARFoundation, a plugin that provides a unified API for working with AR devices, to enable face tracking in AR. ARFoundation uses ARKit on iOS and ARCore on Android to access the built-in face tracking functionality of these platforms [2]. With ARFoundation, developers can create AR apps that track a user's face and use that information to generate realistic 3D effects, such as applying virtual makeup or animating characters to match the user's expressions. Additionally, ARFoundation provides access to the raw data from the face-tracking algorithms, allowing developers to build custom face-tracking experiences tailored to their specific demands. Overall, ARFoundation paves a medium for Unity3D developers to incorporate face tracking into their AR apps, providing a seamless and engaging experience for users. An interpretation of the surface by the camera lens of a mobile device can be shown in figure 1 [3].

Face tracking in Unity can be used to build an Android interface by utilizing the front-facing camera on a mobile device to track the movements and expressions of a user's face. Here's a basic overview of how it can be done: Install a face-tracking plugin or asset from the Unity Asset Store. Configure the camera settings in Unity to use the front-facing camera on a mobile device. Use the face tracking plugin or asset to detect and track the user's face. Map facial expressions or movements to specific actions or interactions in the interface. For example, you can use a smile to trigger an action such as opening a menu or starting a game. Build and run the project on an Android device to test the face-tracking functionality. Fine-tune the facial mapping and adjust the settings as needed to achieve the desired behavior. This is just a basic overview, and the specifics of the implementation will vary depending on the face-tracking plugin or asset used and the specific requirements of the interface being built.

In this paper, we extend well-designed 3D face masks into AR face masks and demonstrate the possibility of transforming this into an NFT-copyrighted AR face mask that is helping authenticate the ownership of the AR mask user. The output of this project will not only potentially validate the value of the NFT technology but also explore how to combine the NFT technology with the AR technology so as to be applied to the e-commerce and e-business aspects of the multimedia industry. More specifically, ARCore plugin has been used to utilize unity, which provides the necessary environment to develop and build an interface for multiple platforms, in this case for an Android device. This Android interface uses Unity's face mesh tracking ability to create a mask that wraps a detected face. The interface is prematurely made to be used with a front-facing interface. As that is the convenient way for a mobile user to use the interface and view the result of the face tracking as they use it.

**Overview**

Figure 2 above shows the flow of states, mentioning the major components of the project. Initiating the first phase, a face-tracking environment in unity3D was created. This is formed by setting a game object in Unity. Preferably, the latest version of unity (2022i.X) is recommended to avoid common Gradle errors reported in previous versions. The Unity documentation does a very good job explaining the dependencies and capabilities of the augmented reality plugin -AR Foundation- [4].

|  (1)  |  (2)  |
|---|---|
|  (3)  |  (4)  |

Figure 5. Show steps to create a new item in an NFT collection.

The second phase is performed when the AR foundation has an environment ready for building. We can then start working on the interface, which will be the medium between the face-tracking augmented reality technology and the end user. To increase engagement and enhance the overall user experience, the ability to switch mask textures was introduced to the app. UI Design: Unity's UI system is used to create interactive user interfaces, such as buttons, text fields, and pop-ups [5]. This was beneficial for the project for giving the ability to add a toggle button to the user's phone screen.

After adding the button, a customized script is used to access the mask toggle feature. C# programming language was used for the script. A function with an efficient algorithm to swap the textures saved in the APK (Android Package Kit). The script containing the function is then linked to the button created and already present in the user interface.

**Technical Approach**

**NFT-Copyright Process.**

NFTs are digital assets that use blockchain technology to verify the ownership and authenticity of a unique item or piece of content. This type of copyright protection can also help prevent unauthorized reproduction or distribution of the AR face mask design [6]. In addition, NFTs are a relatively new concept, and owning an AR face mask with an NFT copyright can help set a precedent for how intellectual property rights are defined and protected in the digital age. However, there may be legal and ethical implications to consider [7], such as the potential for NFT copyright to further entrench existing power imbalances and inequalities in the art world. Overall, the impact of owning an AR face mask with NFT copyright is complex and will likely continue to evolve as the technology and legal frameworks surrounding NFTs develop.

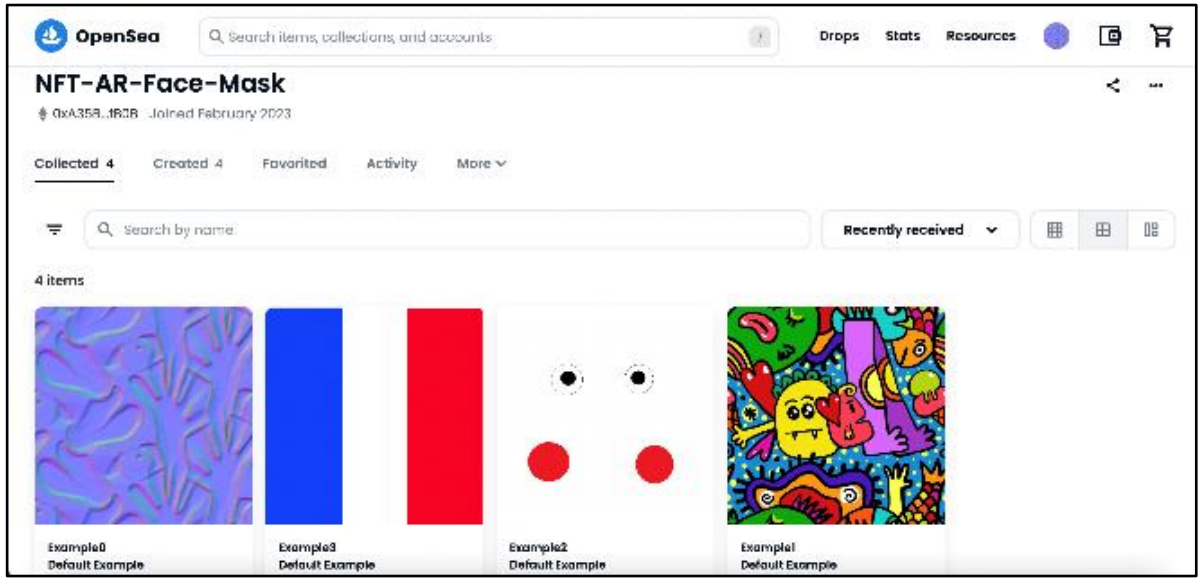

Figure 6. NFT Collection Result

The copyright process for NFTs is different from traditional copyright registration processes. Choosing the appropriate NFT platform to create an AR face mask product can be critical in ensuring the project'sproject's success. There are several factors to consider when selecting an NFT platform, like the type of blockchain and security measures. Thereby, creating an NFT collection with AR face masks in OpenSea involves the following steps, as shown in Figure 5.

1. Choose a Blockchain: Use the Ethereum blockchain because it is one of the most established and widely used blockchain platforms for creating NFTs [8]. Secondly, it has a high level of security and reliability. The Ethereum blockchain uses a consensus mechanism called Proof of Work, which makes it very difficult for anyone to manipulate or alter the blockchain'sblockchain's records. Finally, it is a decentralized platform and provides a secure and tamper-proof way to store these digital assets.

2. Create an NFT: Choose a token with a unique AR face mask, set a name, and determine the size and format [9]. It is helpful to define the parameters for the digital asset and ensure that it meets the desired standards. Each NFT with a unique token ID increases its value and credibility.

3. Create a collection: Set a collection name and upload feature images that reflect the focus of NFTs [10]. Add an "art" tag that combines the uniqueness of NFTs with AR technology's interactive and engaging nature. It brings customization, simplicity, and structure to the NFT collection, and creates a sense of rarity and overlay digital art onto the real world, creating a blended reality experience.

4. Store the NFT: Save each token (AR NFT) to the collection, which result present in Figure 6. It helps to establish a clear and comprehensive framework for the ownership and licensing of a set of AR NFTs. It aslo can protect their rights and ensure that AR NFTs are being used and distributed in a way that aligns with artistic vision.

The NFT copyright process is an important step to protect their intellectual property rights in the emerging digital asset landscape.

**Unity AR Foundation.**

Unity's mobile optimization enhances developers' ability to perform such functions smoothly on mobile devices. Our choice for this project was an Android phone, thus it has only been tested on that platform. The third phase of the project involves building the app after both the face tracking and user interface are set up. For building the Unity app for Android, installing SDKs available with the installation of the unity editor is required. Extended details of this installation for each platform can be found on Unity's website [11].

Components in unity are a medium between the programmer and the development environment. They allow developers to work with pre-built scripts for specific tasks. AR foundation does a great job of offering different tracking depending on what the user wants to be achieved in their interface. For face-tracking, the AR Face Manager component is responsible for that in Unity GameObjects. GameObjects are trackable objects, by unity, to communicate with the lens facilitating surface tracking [12]. This is done once every frame: where faces are added, updated and removed. In this project, the front-facing camera is the most convenient and thus is specified in the component settings. Rear-facing camera also works but since one of the requirements of the project is for the user to be able to look at their face while a mask is generated, the front camera is utilized.

ARCore is one of the contributors to the success of this project. The google-backed application programming interface provides a library of classes for unity development. This is beneficial because it aligns with the resources available to implement the project. Unlike ARKit, to build with ARCore there is no need for a MAC OS X environment. This means we can use any windows OS that can run the unity editor to build this application. With the choice of Google Pixel 7 as the Android mobile device being built to.

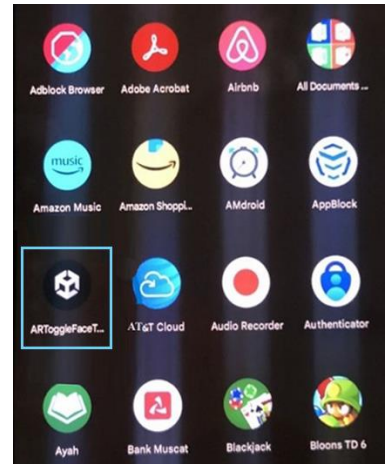

Figure 3: A look at how the APK is displayed in an Android phone.

After setting up Unity with the correct Android SDKs, we can then start building the app based on our needs. Any application needs an interface to be interacted with, this is where we use the Unity toolkit for that. UI toolkits are important for Unity interface development as they provide a comprehensive and streamlined way for developers to create and manage the user interface for their games and applications [5]. A few key benefits of the toolkits in Unity:

1. Consistency: UI toolkits offer pre-built UI elements such as buttons, sliders, text fields, etc. that are consistent with the look and feel of the platform (e.g., Android). This means that developers don't have to spend time creating UI elements from scratch, which saves time and ensures a consistent experience for users.

2. Ease of use: UI toolkits make it easy for developers to create, modify, and manage UI elements without having to write a lot of code. This is because the UI toolkit provides a visual interface that allows developers to drag and drop UI elements onto the screen set their properties, and assign behavior.

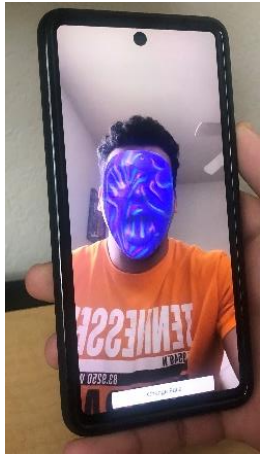

Figure 4: The interface when opening the application.

3. Performance: UI toolkits are optimized for performance, which means that they are designed to run smoothly and efficiently on mobile devices. This is important because mobile devices often have limited resources and processing power compared to desktop computers.

4. Cross-platform compatibility: UI toolkits like the Unity UI Toolkit are designed to work seamlessly across multiple platforms, including Android, iOS, and PC. This means that developers can create UI elements that look and feel the same on different platforms, which makes it easier to maintain and update the UI.

It's important to keep in mind that face tracking can be a complex and resource-intensive task, and it's not always necessary for all types of applications or games. Before choosing a UI system, careful consideration of the requirements of the project and understanding what extent is face tracking involved in it.

Unity's scripting freedom allows customizability in its interfaces. This is introduced by having different texture options to be placed on the surface of the mask. The mask is placed on the user's tracked faces through the AR face manager built into unity and transformed into the mobile device. A key requirement of the interface is to have toggleable mask textures. This is a base feature of an upcoming addition to the application. The addition allows designers to add their own NFT (Non-Fungible Token) textures. This denotes that graphic designers can add the textures they created to the interface through the developer. By making the textures NFT-backed, it will allow these designers to hold the copyrights to their creations and only allow interested users that obtain those NFT textures to use them, exclusively. Any copyright infringement of the textures can be tracked through the serial of the texture stored in the blockchain [13].

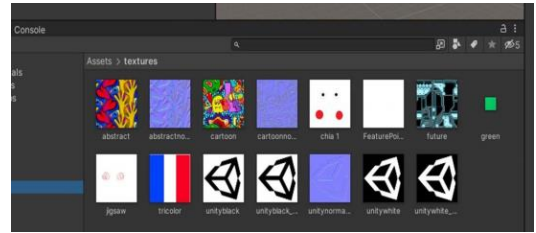

Figure 5: Textures set up for alternation.

**Experimental Results**

Once the interface is built to the Android phone. It can be found as an APK within the phone's apps. It will be shown such as the figure 3.

Then we are ready to view the output of the build. Once the app is opened, we are introduced to a view of the front-facing camera. With a button placed at the bottom of the screen to "change face". The interface does not need any manipulation after it has been built into the phone. This means that face-tracking works automatically. The mask is placed on the face with the first texture in the collection.

The collection of textures in Unity can be seen in the snippet of figure 5. The array holds the textures that are then rotated by the swap function. If more

```
public void SwitchFace()
{
    foreach (ARFace face in faceManager.trackables)
    {
        face.GetComponent<Renderer>().material = faceMaterials[faceMaterialIndex];
    }
    faceMaterialIndex++;

    if(faceMaterialIndex == faceMaterials.Count)
    {
        faceMaterialIndex = 0;
    }
}
```

Figure 6: Snippet of "swap.cs" script showing the "Switchface" function and how it works.

textures are desired then they can be added to the textures file, transformed into materials and associated with the AR face manager component. The swap function is one of the key elements to the customizability of mask textures in the application. Figure 6 is a snippet showing the function unity script. Responsible for the toggling mechanism invoked through the "change face" button.

As mentioned before, this application is based on the ability to add more textures freely. Once a graphic designer wants one of their textures to be in the application, they can add it as a PNG image. Then the process is straightforward from there. Figure 7 shows more examples of textures added for trial.

As can be seen in the snippets of the application. AR foundation does a great job of placing the face mesh on the user's face. Different movements such as eye and mouth movements are tracked smoothly, same can be noticed when moving the head from side to side.

**Conclusion**

In this paper, we propose a 3D modeling approach for NFT-copyrighted AR face masks to improve creative control, brand identification, and ID protection. This project was a valid success and a source of useful insight into the augmented reality sector. The interface brings a systematic approach to the implementation of face tracking in Unity. As well as exploring the ability of custom integration of ownable and copyrighted NFT textures. This emphasizes the ease of the integration of mask textures. Solely to the fact that this interface is now fully built to support such a feature. It has also been tested to be able to diverge these textures given the correct formatting constraints.

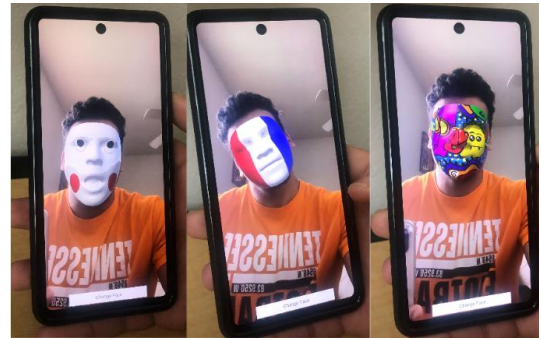

Figure 7: More examples of textures toggled.

As much as we would like to celebrate the achievements of this project. There was a feature that was aimed for but not perfected. It is the existence of hollow eye and mouth sockets. This would have allowed the user to be able to see their eyes and mouth moving and the mask just covering the borders of it. This feature has not been implemented strictly due to the lack of support from Google's ARCore [14]. On the contrary, it is possible to replicate the interface and build the application for an iOS mobile device. Although, this requires as MAC OS X device to run unity editor and Xcode to build the app to an Apple device, it will pave a way for this feature [15]. Of course, that if such a feature is required by the developer or stakeholders, which are the users of the interface in this case.

The world of augmented reality can use the advantages of NFT to reserve the copyrights of its digital output. As more developers get their hands on AR technology, there is a fear that the resultants of the technology are not protected. With the lack of security, users and even developers might be reluctant to use the technology for personal and commercial purposes. One might ask how does storing digital content in the blockchain make it secure. That is done by the addition of uniqueness and ownership [16]. Holders of the original content can spot infringement by owning a unique serial identification that shows proof of work or proof of stake.